\documentclass[useAMS,usenatbib,A4]{mn2e}
\usepackage{graphicx}
\usepackage{natbib}

\def\gro{GRO~J1655-40~}
\def\aap{A\&A~}
\def\mnras{MNRAS~}

\title[Hypersoft state in GRO~J1655-40]{The remarkable timing properties of a `hypersoft' state in GRO J1655-40}
\author[P. Uttley and M. Klein-Wolt]{Philip Uttley$^{1}$\thanks{E-mail:
p.uttley@uva.nl} and Marc Klein-Wolt$^{2,3}$\\
$^{1}$Anton Pannekoek Institute, University of Amsterdam, Science Park 904, 1098 XH Amsterdam, The Netherlands\\
$^{2}$Department of Astrophysics, Research Institute of Mathematics, Astrophysics and Particle Physics, Radboud University Nijmegen,\\
Heijendaalseweg 135, 6525 AJ Nijmegen, The Netherlands\\
$^{3}$Science \& Technology, Olof Palmestraat 14, 2616 LR Delft, The Netherlands
}

\begin{document}

\date{Accepted 2015 April 30.  Received 2015 April 30; in original form 2015 February 4}

\pagerange{\pageref{firstpage}--\pageref{lastpage}} \pubyear{2013}

\maketitle

\label{firstpage}

\begin{abstract}
We report the identification and study of an unusual soft state of the black hole low-mass X-ray binary GRO J1655-40, observed during its 2005 outburst by the {\it Rossi X-ray Timing Explorer}.   {\it Chandra} X-ray grating observations have revealed a high mass-outflow accretion disc wind in this state, and we show that the broadband X-ray spectrum is remarkably similar to that observed in the so-called `hypersoft' state of the high mass X-ray binary Cyg~X-3, which possesses a strong stellar wind from a Wolf-Rayet secondary.  The power-spectral density (PSD) of GRO~J1655-40 shows a bending power-law shape, similar to that of canonical soft states albeit with larger fractional rms.  However, the characteristic bend-frequency of the PSD is strongly correlated with the X-ray flux, such that the bend-frequency increases by two decades for less than a factor 2 increase in flux.  The strong evolution of PSD bend-frequency for very little change in flux or X-ray spectral shape seems to rule out the suppression of high-frequency variability by scattering in the wind as the origin of the PSD bend.  Instead, we suggest that the PSD shape is intrinsic to the variability process and may be linked to the evolution of the scale-height in a slim disc.  An alternative possibility is that variability is introduced by variable absorption and scattering in the wind.  We further argue that the hypersoft state in GRO~J1655-40 and Cyg~X-3 is associated with accretion close to or above the Eddington limit.
\end{abstract}


\section{Introduction}
Galactic black hole X-ray transients are useful laboratories for studying the accretion of matter in the strongest gravitational fields in the Universe.  During the course of an outburst, an accreting black hole transient undergoes a variety of distinct spectral states marked by differing contributions from the `standard' optically thick accretion disc (which emits as a multi-temperature blackbody) and the more mysterious power-law emission, which may originate from an optically thin coronal region (e.g. \citealt{Done2007}).  These different states clearly imply large changes in the structure of the innermost accretion flow, where most of the energy is released.  

The spectral changes are also accompanied by strong, correlated changes in the X-ray variability (or `timing') properties (e.g. \citealt{Homan2001,Belloni2005,Remillard2006}).  `Hard' (power-law dominated) states show the most (few tens of per cent rms) variability, across a relatively broad range of time-scales.  The transitional or `intermediate' states show the appearance of strong quasi-periodic oscillations (from $\sim0.1-10$~Hz) as well as a concentration of the broader-band noise, with variability becoming concentrated towards higher temporal frequencies as the spectrum softens, due to the evolution in frequency and strength of the different power-spectral components (e.g. \citealt{Klein-Wolt2008}).  Finally, in the `soft' state where disc blackbody emission dominates, variability amplitudes are very low (often $<1$ per cent in fractional rms) although the variability is once again spread over a broader range of frequencies, as in the hard state (e.g. \citealt{Heil2014}).  Besides the strong correlation with spectral state of general timing properties such as Power Spectral Density (PSD) shape and amplitude, it is also clear from many observations to date that the characteristic frequencies are typically strongly correlated with spectral shape, and less so with total luminosity (e.g. \citealt{Sobczak2000}).

Besides the general relevance for understanding the behaviour of accreting matter in strong-field gravity, the evolution of outbursting black hole transients has developed a wider astrophysical significance in highlighting the connection of mechanical outflows to the evolution of the source.  It is now well-established that the presence and power of radio-emitting jets is strongly linked to spectral state (e.g. \citealt{Fender1999,Corbel2001,Klein-Wolt2002,Miller-Jones2012}).  A recent twist to the story has been the discovery that high binary-orbit inclination black hole transients in their soft states show the presence of outflows, inferred from X-ray absorption lines (e.g. \citealt{Miller2006a,Miller2006b,Ueda2009,Ponti2012}).  The natural implication is that these outflows are associated with equatorial disc winds, rather than in a collimated jet-type flow. 

The most powerful (possibly magnetically-driven) disc wind discovered in a black hole X-ray transient to date, was observed in the low mass X-ray binary and Galactic black hole candidate, GRO~J1655-40, during its 2005 outburst \citep{Miller2006b,Miller2008,Neilsen2012}.  This long outburst is interesting for a number of reasons (e.g. see \citealt{Motta2012}), not least due to the exceptional coverage of the outburst by an extensive monitoring campaign with the Rossi X-ray Timing Explorer ({\it RXTE}), which provided daily observations of up to 10~ks per day.  In this paper, we use a part of the intense GRO~J1655-40 2005 {\it RXTE} monitoring campaign to reveal a hitherto unknown aspect of black hole X-ray transient timing behaviour, which seems to be linked to the unusual spectral state that produces the strong wind in this system.

\section{Observations, data reduction and analysis}
\label{observations}
\gro was observed intensively by the {\it Rossi X-ray Timing Explorer} ({\it RXTE}) throughout its 2005 outburst, providing a comprehensive data set for X-ray spectral and timing analysis.  Here we use data from two of the data modes of {\it RXTE}'s Proportional Counter Array (PCA), specifically the `Standard 2' mode data which provides optimal spectral resolution but low (16~s) time resolution and Single Bit mode data, which provides high (122~$\mu$s) time resolution but in only two energy bands, corresponding to $\sim2-6$~keV and $\sim6-15$~keV (which we refer to here as the soft and hard bands respectively).  

\begin{figure}
\includegraphics[scale=0.33]{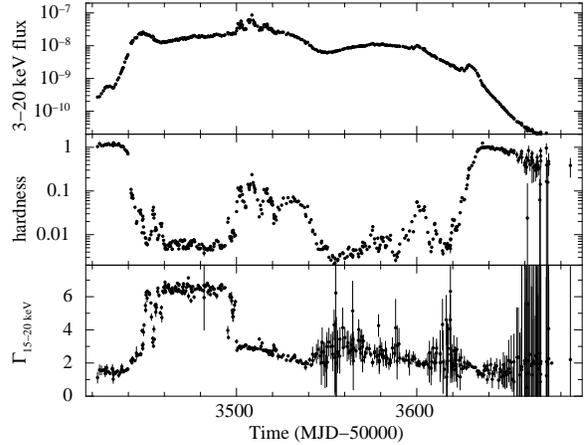}
\caption{The 2005 outburst of \gro.  The top panel shows the 3-20 keV X-ray flux (units are erg~cm$^{-2}$~s$^{-1}$), the middle panel shows the hardness, which is defined as the ratio of 15-20~keV flux to 3-5~keV flux, while the bottom panel shows the photon index measured by fitting spectra in the 15-20~keV band.  See Section~\ref{observations} for further details of how these values are measured.}
\label{lcurves}
\end{figure}

We extracted source spectra from the Standard 2 PCA data, and generated model background spectra and response matrices using the standard recipes within the {\sc heasoft v6.12} software package.  The spectra and response matrices were generated separately for each {\it RXTE} ObsID (typically corresponding to between 1 and 3 satellite orbits), obtained throughout the entire outburst.  Since the number of PCA proportional counter units (PCUs) switched on is variable, we extracted spectra only for PCU2, which was always on.  Fluxes are generated in a model-independent way by dividing the spectrum by the effective area of the instrument response in each spectral channel.  This is carried out by unfolding the spectrum with respect to a zero-slope power-law (i.e. a constant) in the {\sc xspec} spectral-fitting software, and measuring the unfolded flux over the specified energy range (interpolating where the specified energy does fall neatly at the edge of a spectral channel).  Here we obtained 3-20~keV fluxes for each observation, as well as fluxes in the 3--5~keV and 15--20~keV energy bands, in order to measure a hardness ratio.  The resulting outburst light curve and hardness ratios are plotted in the top two panels of Fig.~\ref{lcurves}.

To obtain PSDs, we use the standard recipe to extract light curves from the Single Bit data and our own in-house code to make PSDs.  The PSDs are fitted using the {\sc xspec} software, by using the {\sc flx2xsp} {\sc ftool} within {\sc heasoft} to convert the PSDs into the {\sc xspec}-readable PHA spectral file format and create diagonal response matrices.

\section{Identifying an unusual soft state in the 2005 outburst}
\label{manxid}
\begin{figure}
\includegraphics[scale=0.33]{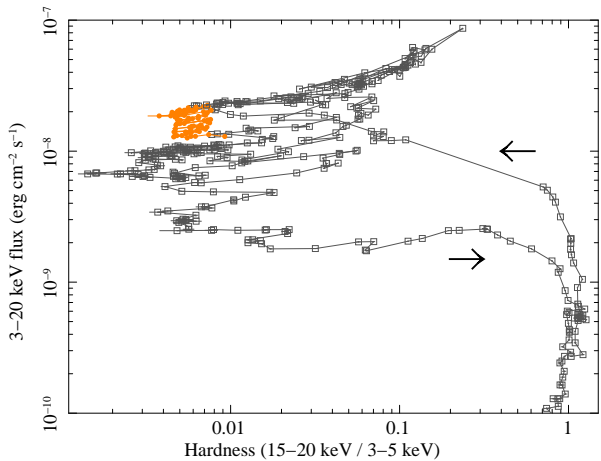}
\caption{Hardness-intensity diagram for the 2005 outburst of \gro.  The lines connect observations that are contiguous in time and arrows indicate the evolution of the source along the hard-soft and soft-hard state transitions at the beginning and end of the outburst respectively.  Filled orange circles mark the stable hypersoft state (see text in Section~\ref{manxid} for definition), and open squares mark the data obtained at other times.}
\label{hardint}
\end{figure}

The 2005 outburst of \gro\ lasted for more than 200 days, and shows a complex pattern of behaviour.  As can be seen in Fig.~\ref{lcurves}, the evolution of the hardness ratio shows evidence for two epochs when the source was in a generally spectrally-soft state (roughly from MJD~53450--53500 and then from MJD~53540--53620), separated by a period with a more intermediate hardness ratio.  The hardness-intensity diagram for the outburst is shown in Fig.~\ref{hardint}.  The source follows a `q'-shaped track that is similar to the hardness-intensity tracks followed by other black hole X-ray binary systems in outburst (e.g. \citealt{Homan2001,Belloni2005,Dunn2010}).  Comparison with these other sources identifies the two epochs with low hardness with the soft state of black hole X-ray binaries (BHXRBs). A transition to an intermediate state separates these two epochs.  However, a closer look at the spectral shapes corresponding to these two soft-state epochs reveals some interesting differences between them.

\begin{figure}
\includegraphics[scale=0.33]{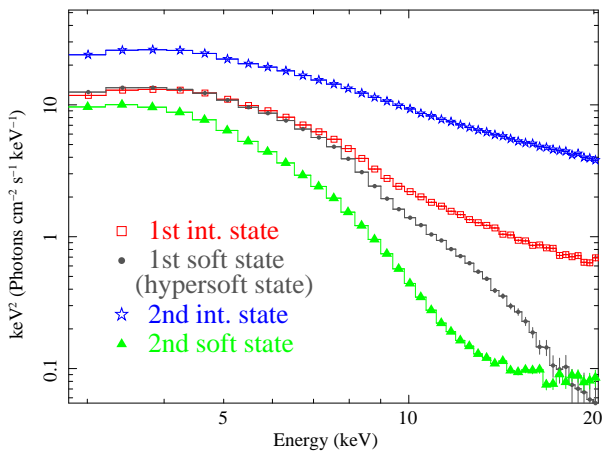}
\caption{Comparison of intermediate and soft state spectra from the 2005 outburst of \gro. Note that the spectra have been unfolded relative to a zero-slope power-law model to account for the instrumental effective area but maintain instrumental resolution (see Section~\ref{observations} for details).  The {\it RXTE} Observation IDs (and corresponding dates) shown are as follows: red open squares (first intermediate state): 91702-01-12-01 (MJD~53453); dark grey filled circles (first soft state, classified as the `hypersoft' state in this paper): 91702-01-30-04 (MJD~53473); blue open stars (second intermediate state): 91702-01-53-04 (MJD~53501); green filled triangles (second soft state): 91702-01-15-10 (MJD~53573).}
\label{speccomp}
\end{figure}

In Fig.~\ref{speccomp} we plot four examples of spectra of \gro\ covering the 3--20~keV range, comparing the spectra obtained in periods of intermediate hardness just before the first soft state and in between the two soft states, with the spectra from each soft state.  The intermediate-hardness spectra are fairly typical of spectra from the `canonical' intermediate states of other BHXRBs, showing relatively strong, steep power-law tails, probably also combined with thermal emission from the accretion disc.  The spectrum from the second soft state shows a canonical soft state spectrum, with a strong disc thermal component together with a weak power-law tail, which flattens the spectrum above $\sim 12$~keV.  However, the spectrum from the first soft state is quite distinct from that of the second soft state, in that it shows an unusual spectrum with either a very steep or even absent power-law tail, with no obvious flattening above 12~keV.  This spectrum appears quite different to the canonical soft state spectra seen in the second soft-state epoch and in other BHXRBs, where there is always a weak power-law tail in addition to the thermal blackbody emission (e.g. \citealt{Done2007}).  

\begin{figure}
\includegraphics[scale=0.33]{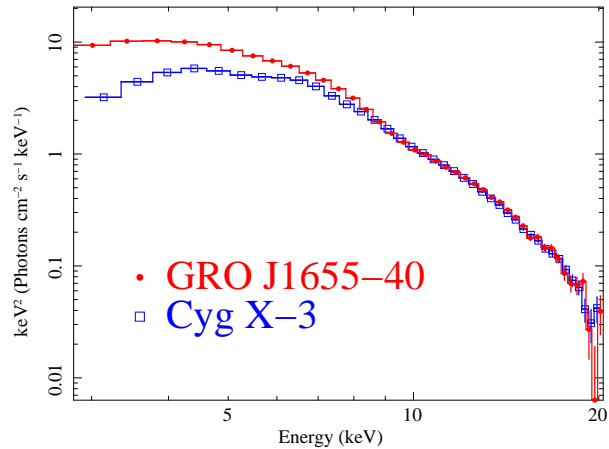}
\caption{Comparison of unfolded spectra from the unusual soft state of \gro (ObsID 91702-01-21-00) and the hypersoft state of Cygnus~X-3 (50062-02-03-02).}
\label{cygx3comp}
\end{figure}

A similar spectral state has been noted and classified as the `hypersoft state' by \citet{Koljonen2010} in the high mass X-ray binary Cyg X-3.  In Fig.~\ref{cygx3comp} we show an example {\it RXTE} PCA spectrum of the hypersoft state of Cyg X-3 for comparison with that of \gro in its unusual soft state.  The \gro spectrum is chosen so that the flux above 10 keV matches that seen in Cyg X-3, but is similar in shape to other spectra from this state of GRO~J1655-40.  Above 10~keV the two spectra are remarkably similar in shape as well as flux.  Below 10~keV, the main differences may be due to the larger neutral absorbing column (few$\times 10^{22}$~cm$^{-2}$ versus $7.4\times 10^{21}$~cm$^{-2}$) that is required to fit the spectrum in Cyg~X-3 \citep{Koljonen2010}.  Considering the close similarity between the spectra above 10~keV, we will classify the unusual soft state of \gro as `hypersoft' for the remainder of this paper.

We also compared data from the {\it RXTE} {\it High Energy X-ray Timing Experiment} ({\it HEXTE}) instrument for the same observations of Cyg~X-3 and GRO~J1655-40 shown in Fig.~\ref{cygx3comp}.  The {\it HEXTE} spectra are consistent with zero flux above 30~keV at the 90~per~cent confidence level, as would be expected from extrapolation of the steep spectral shape seen at lower energies.  We do not consider the {\it HEXTE} data further in this study.

The hypersoft and canonical soft states have similar low values of hardness ratio (e.g. see Figs.~\ref{lcurves} and \ref{hardint}), but the hypersoft state can be easily distinguished from the canonical soft state by fitting a simple power-law to the 15-20~keV part of the spectrum.  The result of such fits to all the data in the outburst are shown in the bottom panel of Fig.~\ref{lcurves}.  The hypersoft state is clearly defined by the 15-20~keV photon index $\Gamma_{15-20}>5$ (c.f. $\Gamma_{15-20}=2$--3 for the canonical soft state).  At the beginning and end of the hypersoft state there are some large fluctuations in $\Gamma_{15-20}$ where the power-law tail apparently flattens, but during the times from MJD~53459.0 to MJD~53494.0 the values of $\Gamma_{15-20}$ are stable and the hypersoft state appears to persist throughout this entire period.  We highlight the corresponding data points with filled circles in the hardness intensity diagram of Fig.~\ref{hardint}, which shows that the hypersoft state is marked only by subtle differences from the canonical soft state in terms of this measure of hardness, even though the spectrum itself is very different from that of the canonical soft state.

It is especially interesting to note that the {\it Chandra} observation which revealed evidence for strong wind absorption in GRO~J1655-40 \citep{Miller2006b} was obtained on MJD~53461, which is during the hypersoft state.  A previous {\it Chandra} observation, which revealed significantly fewer and weaker absorption features, took place on MJD~53441 and can be identified with the state transition which preceded the hypersoft state.

\section{The unusual timing properties of the hypersoft state in GRO~J1655-40}
\label{timing}
We now consider the timing properties of the hypersoft state in GRO~J1655-40.  Throughout this analysis and in the rest of the paper we use for convenience the definition that the hypersoft state corresponds to observations with  $\Gamma_{15-20}>5$.  This definition is motivated by the relatively stable range of steep $\Gamma_{15-20}$ observed from MJD~53459--53494, but is not meant to be a formal classification.

\subsection{PSD shape}
\label{psds}
\begin{figure}
\includegraphics[scale=0.33]{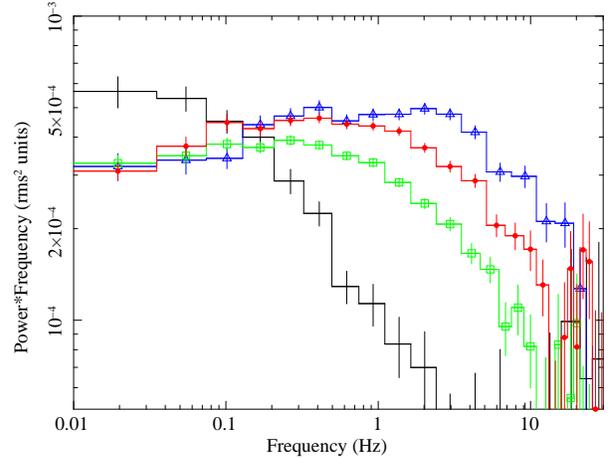}
\caption{Comparison of Poisson-noise-subtracted PSDs within the hypersoft state.  The {\it RXTE} Observation IDs (and corresponding dates) shown are as follows: black crosses: 91702-01-18-00 (MJD~53460); red filled circles: 91702-01-33-00 (MJD~53476); green open squares: 91702-01-34-01 (MJD~53478); blue open triangles: 91702-01-42-00 (MJD~53488).}
\label{indpsdcomp}
\end{figure}

We first examined the PSDs of individual observations in the hypersoft state.  We use the Single Bit mode data in the soft band only, to minimise any distortions to the PSD shape due to energy-dependent effects (we will consider these effects briefly in Section~\ref{specrelationship}).  To examine the PSDs in the absence of their Poisson noise components, we construct PSDs up to 512~Hz and fit only the region above 200~Hz with a single power-law (with free slope and normalisation), in order to model the Poisson noise component.  We then extrapolate this fit to lower frequencies and plot the residuals to reveal the intrinsic source PSDs.  Fig.~\ref{indpsdcomp} shows these PSDs from four observations during the hypersoft state, chosen to cover a range of times and flux levels (see caption for details).  Note that we use the common practice of plotting power$\times$frequency, which reveals the frequencies which dominate the variability, just as plots of energy$\times$flux-density used to plot spectral energy distributions reveal the dominant components contributing to a source's luminosity. The resulting values are plotted in fractional rms-squared units\footnote{Strictly speaking these values correspond to the fractional rms-squared contained in a factor $e$ range in frequency assuming a PSD slope of -1 (a flat top in power$\times$frequency), e.g. a value of $5\times10^{-4}$ corresponds to a fractional rms of roughly 3.4~per~cent over a decade range in frequencies}.  

The PSDs are characterised by broadband noise, with a wide range of frequencies dominating the variability and no evidence for quasi-periodic oscillations (QPOs).  There is, however, a type of characteristic time-scale in the form of a bending cut-off at high-frequencies.  The cut-off bend-frequency increases significantly through the hypersoft state, from $\sim 0.1$~Hz to $\sim 10$~Hz.  Interestingly, the evolution of this bend frequency isn't simply monotonic with time, since the PSD with the second-highest bend frequency in Fig.~\ref{indpsdcomp} (filled-circles) is from an observation obtained two days {\it before} the PSD with the third-highest frequency (open squares).  The key difference between these observations is the source flux, since the second-highest bend-frequency also corresponds to the observation with the second-highest source flux of the four observations shown. For the observations shown, as the bend frequency increases the 3-20~keV flux has ascending values of $(1.3, 1.8, 2.0, 2.2)\times10^{-8}$~erg~cm$^{-2}$~s$^{-1}$.

\begin{figure}
\includegraphics[scale=0.33]{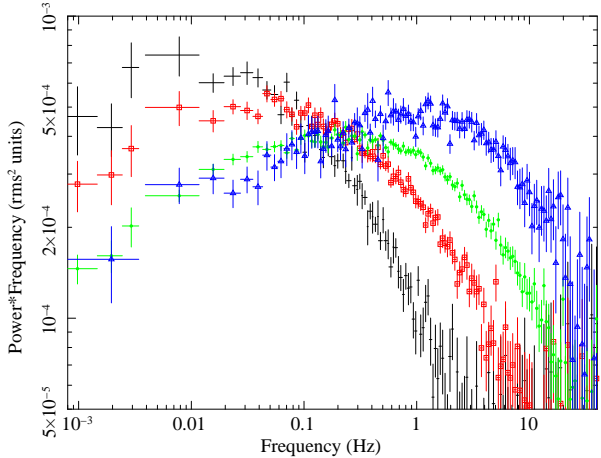}
\caption{Comparison of Poisson-noise-subtracted PSDs made from observations grouped according to flux (see Table~\ref{groupids}.  For clarity PSDs from only four groups are shown: group 1 (black, crosses), group 3 (red, open squares), group 5 (green, filled circles) and group 7 (blue, open triangles).}
\label{groupedpsds}
\end{figure}

Since the PSD appears to evolve strongly with changes in the source flux, it is useful to quantify these changes using higher signal-to-noise PSDs constructed from groups of observations with similar flux.  We therefore assigned the observations of the hypersoft state during the time range MJD 53459--53491 (chosen to span the full range of fluxes observed within the time range where $\Gamma_{15-20}$ is relatively stable) into 7 groups corresponding to roughly equal flux intervals, according to the average flux measured in each observation.  The Observation IDs corresponding to each group are listed in Table~\ref{groupids}.  We produced PSDs from the combined light curves obtained from the Single Bit mode soft band data.  To constrain the broadband PSD shape it is useful to extend the coverage of the PSDs to lower frequencies, so we generated PSDs using 1024~s duration light curve segments in addition to the standard 128~s duration segments.  This produced PSDs with slightly lower signal-to-noise than for 128~s segments, since 1024~s segments do not efficiently cover the single-orbit contiguous {\it RXTE} exposures.  Therefore we only used the low-frequency parts of these PSDs, below 0.0078~Hz, in combination with the PSDs made from 128~s segments to cover higher frequencies.  The PSDs were binned up in frequency over a minimum geometric spacing of 1.05.  Fig.~\ref{groupedpsds} shows the residual intrinsic source PSDs, over and above the Poisson noise level, obtained in the same way as for Fig.~\ref{indpsdcomp}.  The PSDs from all groups show similar evolution with flux, but for clarity we show only the PSDs for groups 1, 3, 5 and 7.  The PSD bend frequency clearly evolves strongly towards higher frequencies as flux increases from group 1 to group 7.
\begin{table}
\centering
\caption{Observation dates and {\it RXTE} Observation IDs assigned to each group.  For conciseness the format is date:ID, where the date is in MJD-53000 and the full Observation IDs are of the form 91702-01-ID.  The flux ranges covered by the observations in each group are given for the 3-20~keV band.}
\label{groupids}
\begin{tabular}{@{}cccc@{}}
\hline
 Group 1 : & \multicolumn{3}{c}{$(1.26-1.38)\times 10^{-8}$~erg~cm$^{-2}$~s$^{-1}$} \\
\hline
459:17-01 & 459:17-03 & 460:18-00G & 460:18-01 \\ 
460:18-02 & 460:18-03 & 460:18-04 & 461:19-00 \\
464:22-00 & & & \\
\hline
 Group 2 : & \multicolumn{3}{c}{$(1.38-1.52)\times 10^{-8}$~erg~cm$^{-2}$~s$^{-1}$} \\
\hline
463:19-01 & 462:20-00 & 462:20-01 & 462:20-02 \\
463:21-00 & 463:21-01 & 464:22-02 & 464:22-03 \\
465:23-00 & 466:24-02 & 467:25-02 & 467:25-03 \\
\hline
 Group 3 : & \multicolumn{3}{c}{$(1.52-1.66)\times 10^{-8}$~erg~cm$^{-2}$~s$^{-1}$} \\
\hline
466:24-00 & 466:24-03 & 467:25-00 & 467:25-01 \\
468:26-00 & 468:26-01 & 468:26-02 & 469:27-00 \\
470:27-02 & 470:28-01 & 470:28-02 & 472:29-02 \\
471:30-00 & & & \\
\hline
 Group 4 : & \multicolumn{3}{c}{$(1.66-1.80)\times 10^{-8}$~erg~cm$^{-2}$~s$^{-1}$} \\
\hline
469:27-01 & 470:28-00 & 471:29-00 & 471:29-01 \\
472:30-01 & 472:30-02 & 473:30-03 & 473:30-04 \\
474:31-00 & 473:31-01 & 473:31-02 & 478:34-01 \\
\hline
 Group 5 : & \multicolumn{3}{c}{$(1.80-1.94)\times 10^{-8}$~erg~cm$^{-2}$~s$^{-1}$} \\
\hline
474:31-03 & 474:31-04 & 475:32-00 & 475:32-01 \\
475:32-02 & 477:34-00 & 479:35-00 & 479:35-01 \\
479:35-02 & 481:36-00 & 481:36-03 & 482:37-00 \\
482:37-01 & 483:38-00 & 484:38-02 & 484:39-00 \\
485:40-02 & 485:40-03 & 486:41-01 & \\
\hline
 Group 6 : & \multicolumn{3}{c}{$(1.94-2.08)\times 10^{-8}$~erg~cm$^{-2}$~s$^{-1}$} \\
\hline
476:33-00 & 476:33-01 & 479:34-02 & 480:36-01 \\
487:38-01 & 484:38-03 & 485:40-00 & 487:41-00 \\
490:44-00 & 490:44-03 & 491:45-00 & 491:45-01 \\
\hline
 Group 7 : & \multicolumn{3}{c}{$(2.08-2.24)\times 10^{-8}$~erg~cm$^{-2}$~s$^{-1}$} \\
\hline
487:41-02 & 488:42-00 & 489:43-00 & 489:43-01 \\
491:44-01 & & & \\
\hline
\end{tabular}
\end{table}

To quantify the evolution of the PSDs through the different flux groups, we fitted them with a bending power-law model of the form:
\begin{equation}
P(\nu)=\frac{A \nu^{-\alpha_{\rm L}}}{1+(\frac{\nu}{\nu_{\rm b}})^{(\alpha _{\rm H}-\alpha _{\rm L})}}
\end{equation}
where $P(\nu)$ is the power at frequency $\nu$, $A$ is a normalising factor, $\nu_{\rm b}$ is a bend frequency and $\alpha_{\rm L}$ and $\alpha_{\rm H}$ are the low and high-fequency power-law slopes, below and above the bend respectively.  \citet{McHardy2004} found that such a function provides a good fit to the PSD of the BHXRB Cyg~X-1 in its soft state, fitting significantly better than a sharply broken power-law.  The bending power-law also appears to be a suitable function to mimic the smooth bend seen in the hypersoft state PSDs of GRO~J1655-40, which show no strong evidence for sharper features such as QPOs.  We fitted the data over the whole frequency range from ~1~mHz to 512~Hz, including a power-law with free slope and normalisation to fit the Poisson noise component which dominates at frequencies above 10-100~Hz.

Inspection of the PSDs in Fig.~\ref{groupedpsds} shows that both the bend frequency and normalisation of the PSDs vary significantly.  However, it isn't clear whether the low and high-frequency slopes also change, especially because the different PSDs are sensitive to the low and high-frequency slopes to different degrees, as the bend-frequency evolves (e.g. low bend frequencies give a stronger constraint on high-frequency slope and vice versa).  Therefore, we first fitted all seven grouped PSDs together with the bending power-law model where the bend frequencies and normalisations were left free but the low and high-frequency slopes were each tied to be the same in all PSDs.  

The best-fitting model returned slopes $\alpha_{\rm L}=0.83\pm0.01$ and $\alpha_{\rm H}=1.77\pm0.01$ for a $\chi^{2}$ of 1392 for 1256 degrees of freedom (d.o.f.).  Examination of the residuals suggested that the fit was worse at higher frequencies so that untying the high-frequency slopes to vary between groups may improve the fit. Keeping $\alpha_{\rm L}$ tied to be the same for all groups, but freeing $\alpha_{\rm H}$, we found that the fit improved significantly to $\chi^{2}=1350$ for 1250 d.o.f., or a $\Delta \chi^{2}$ of 42 for 6 additional free parameters, which according to an F-test is a significant improvement at the 99.9999\% ($4.9\sigma$) confidence level.  The corresponding fixed value of low-frequency slope is $\alpha_{\rm L}=0.82\pm0.01$.  We next untied the low-frequency slopes so that all parameters were free to vary.  The resulting improvement was slight, with $\Delta \chi^{2}=13.6$ for 6 additional free parameters, which according to the F-test is only marginally significant at the 95\% ($<2\sigma$) confidence level.  We therefore conclude that the PSDs are consistent with evolution in frequency, normalisation and high-frequency slope, but possess a constant low-frequency slope, $\alpha_{\rm L}=0.82$.  The best-fitting parameters for this model are given in Table~\ref{psdparams}, together with the average fluxes and hardness ratios for the corresponding grouped flux spectra.
\begin{table}
\centering
\begin{minipage}{90mm}
\caption{Results of fits to PSDs of grouped data, showing parameters of the best-fitting bending power-law model with low-frequency slope tied to find the same value ($\alpha_{\rm L}=0.82\pm0.01$) for all groups.  Note that
the relative uncertainties in flux and hardness are less than 0.01\% and 1\% respectively.}
\label{psdparams}
\begin{tabular}{@{}lccccc@{}}
\hline
Group & Flux\footnote{3-20~keV, in units of $10^{-8}$~erg~cm$^{-2}$~s$^{-1}$.} & Hardness\footnote{(15-20 keV)/(3-5 keV) flux ratio.} & 
$\nu_{\rm b}$ & $\alpha_{\rm H}$ & $A_{1}$\footnote{Normalisation in units of $10^{-3}$~Hz$^{-1}$ defined as the rms$^{2}$ power at 1~Hz obtained from extrapolating the low-frequency power-law to that frequency.}\\
\hline
1 & 1.31 & 0.0052 & $0.09\pm0.02$ & $1.87\pm0.03$ & $1.44^{+0.23}_{-0.11}$ \\ 
2 & 1.44 & 0.0062 & $0.15^{+0.02}_{-0.01}$ & $1.79\pm0.03$ & $1.14^{+0.09}_{-0.08}$ \\
3 & 1.59 & 0.0061 & $0.26^{+0.02}_{-0.04}$ & $1.70\pm0.03$ & $1.03^{+0.04}_{-0.07}$ \\
4 & 1.76 & 0.0062 & $0.69^{+0.09}_{-0.05}$ & $1.72\pm0.04$ & $0.74^{+0.04}_{-0.04}$ \\
5 & 1.88 & 0.0052 & $1.26^{+0.14}_{-0.07}$ & $1.72\pm0.04$ & $0.63^{+0.02}_{-0.03}$ \\
6 & 2.02 & 0.0058 & $2.30^{+0.27}_{-0.23}$ & $1.80\pm0.06$ & $0.56^{+0.02}_{-0.02}$ \\
7 & 2.19 & 0.0068 & $4.89^{+0.61}_{-0.48}$ & $1.92\pm0.11$ & $0.55^{+0.02}_{-0.02}$ \\
\hline
\end{tabular}
\end{minipage}
\end{table}

\begin{figure}
\includegraphics[scale=0.33]{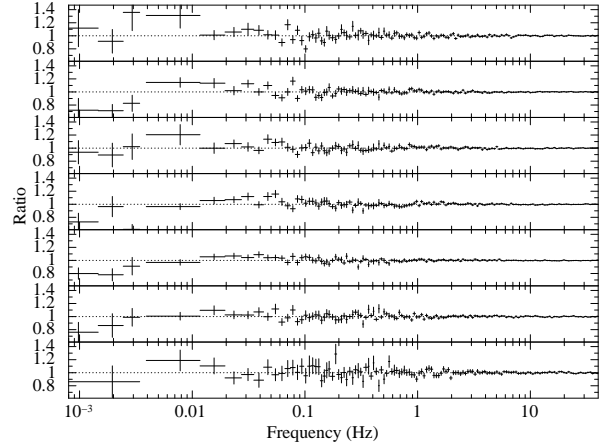}
\caption{Data/model ratios for the best-fitting bending-power-law models described in Table~\ref{psdparams}.  From top to bottom the panels show the ratios for the fits from group 1 to group 7.}
\label{psdratios}
\end{figure}

Note that, although we have fitted the PSDs together in order to constrain the low-frequency slope, no one group's PSD shows any clear deviations from this model, as can be seen in Fig.~\ref{psdratios}, which shows the ratios of the data to the tied $\alpha_{L}$ model for all the groups.  Any deviations from the model are weak, and no clear features can be seen.  

\subsection{Time-evolution of the PSD}
\label{psdvstime}
To study the evolution of the PSD during the hypersoft state in more detail, we next fitted the bending power-law model to the individual observations in the time range MJD~53441--53499, which incorporates all the hypersoft state data ($\Gamma_{15-20}>5$) as well as some `non-hypersoft' data with $\Gamma_{15-20}\leq5$.  Since the grouped data show that the PSDs in the hypersoft state are well-fitted with a bending power-law with constant low-frequency slope, we froze the low-frequency slope at the best-fitting value ($\alpha_{\rm L}=0.82$), in order to better constrain the fit parameters for individual observations, which are necessarily of lower signal-to-noise than the grouped data.  We then proceeded to fit the bending power-law (including additional Poisson noise power-law) to the individual observations, recording the best-fitting values of bend frequency $\nu_{\rm b}$, high-frequency slope $\alpha_{\rm H}$, normalisation $A_{1}$ and also the acceptance probability of the fit, $P_{\rm accept}$.  

\begin{figure*}
\includegraphics[scale=0.6]{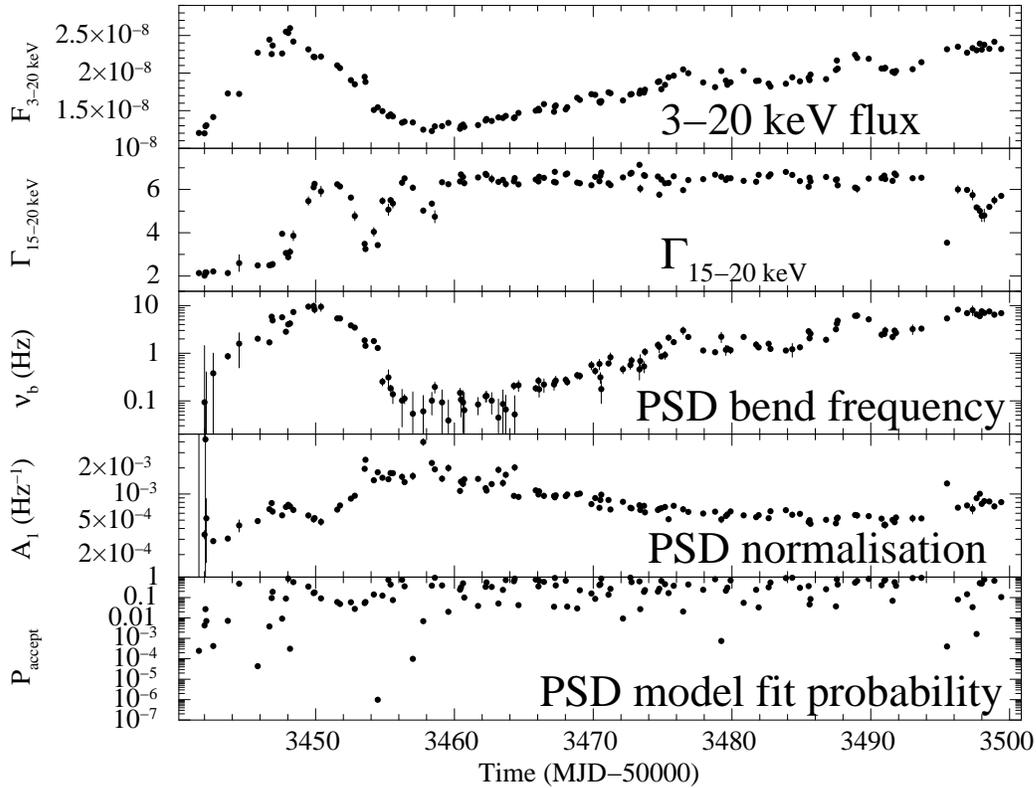}
\caption{Time variation of flux and PSD parameters during and surrounding the hypersoft state.  See Section~\ref{psdvstime} for details.}
\label{psdparamsvstime}
\end{figure*}

The time dependence of the fitted PSD parameters is shown in Fig.~\ref{psdparamsvstime}, together with the 3-20~keV flux and $\Gamma_{15-20}$ for comparison.  Errors on the high-frequency slope $\alpha_{\rm H}$ were fairly large and therefore not informative, so these values are not included in the plot.  It is clear that for the non-hypersoft data the model fits are usually poor, with low acceptance-probabilities.  Inspection of the poorly-fitted PSDs shows an excess of low-frequency power, although substantial improvements in the fits cannot be made by allowing the low-frequency slope $\alpha_{\rm L}$ in the bending power-law model to be a free parameter.  Additional components seem to be required to adequately fit these PSDs, unlike those within the hypersoft state itself, but we do not investigate these non-hypersoft state PSD shapes in more detail here.   

The data show evidence for a clear and systematic dependence of the bend frequency on the flux.  This correlation between bend-frequency and flux is clearest during the hypersoft state.  The correlation applies over a broad range of time-scales, i.e. bend-frequency variations correlated with day-scale changes in flux can be seen, in addition to longer-term changes.  The PSD~normalisation shows a long-term decreasing trend in the hypersoft state.

\subsection{Flux-evolution of the PSD bend frequency}
\label{psdvsflux}
\begin{figure}
\includegraphics[scale=0.33]{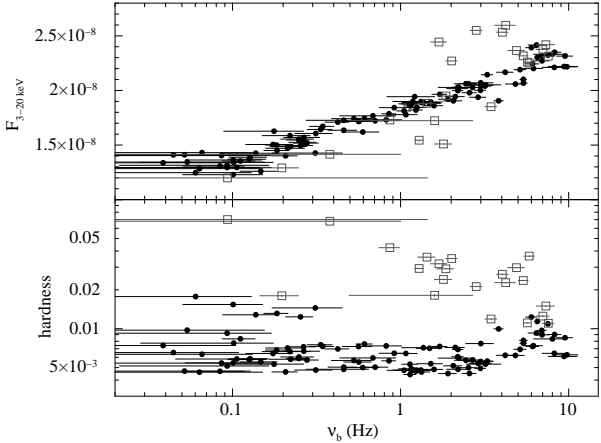}
\caption{Comparison of PSD bend frequency with 3-20 keV flux and 15-20~keV$/$3-5~keV spectral hardness for the hypersoft ($\Gamma_{15-20} >5$, filled circles) and non-hypersoft (open squares) data for \gro in the time range MJD~53441--53499.  Note that four non-hypersoft data points are outside the plot range, consistent with their large errors on bend frequency.}
\label{freqvsfluxcol}
\end{figure}

In Fig.~\ref{freqvsfluxcol} we plot the dependence of the bend-frequency on 3-20~keV flux and spectral hardness, using the data shown in Fig.~\ref{psdparamsvstime}.  The points from the hypersoft observations ($\Gamma_{15-20} >5$) are marked with filled circles, while the other data points are marked with open squares.  The strong and tight correlation of bend-frequency with flux is very striking in the hypersoft state, while there is no correlation between bend-frequency and hardness, despite a factor $>50$ change in bend-frequency.  The non-hypersoft data shows a much weaker correlation of bend-frequency with flux.  As expected, the spectra are also systematically harder than the hypersoft state spectra at the same bend frequencies.

The relationship between flux ($F$, expressed in units of $10^{-8}$~erg~s$^{-1}$~cm$^{-2}$) and bend-frequency ($\nu_{\rm b}$) for the hypersoft state data can be moderately well-fitted with either an exponential, $\nu_{b}=a\exp(cF)$, or a power-law $\nu_{b}=kF^{\alpha}$, with deviations caused by random scatter of about 0.15 dex in frequency, rather than any systematic mismatch of these functions to the data.  For the exponential model, the best-fitting $\chi^{2}=151$ for 77 d.o.f., with best-fitting parameters $a=(2.1\pm0.6)\times10^{-4}$ and $c=4.62\pm0.13$ (all errors are for $\Delta \chi^{2}=2.7$).  For the power-law model, the best-fitting $\chi^{2}=146$ for 77 d.o.f., with best-fitting parameters $k=(6\pm1)\times 10^{-3}$, $\alpha = 8.53\pm 0.27$.  The fact that the data can be fitted with either an exponential or a steep power-law dependence of frequency on flux reflects the very strong dependence of bend-frequency on flux, combined with the relatively narrow flux range the relationship is measured over.

\subsection{The X-ray spectrum and energy-dependence of the PSD}
\label{specrelationship}
\begin{figure}
\includegraphics[scale=0.33]{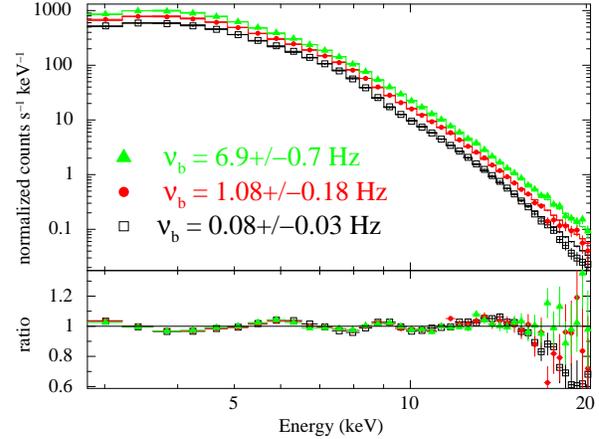}
\caption{Comparison of spectra and data/model ratios obtained during the hypersoft state of GRO~J1655-40, fitted with the simple absorbed disc blackbody plus scattered power-law spectral model described in the text.  The spectra correspond to three observations, ObsIDs 91702-01-19-00 (black, open squares) 91702-01-31-02 (red, filled circles), 91702-01-49-02 (green, filled triangles), the PSDs of which span a wide range of bend frequencies, listed on the figure.  The spectrum with the lowest bend-frequency was obtained during the {\it Chandra} observation which revealed a strong disc wind.}
\label{hypspeccomp}
\end{figure}

In Fig.~\ref{hypspeccomp} we show X-ray spectra (in units of detector counts per keV) from three example observations of the hypersoft state which cover a wide range of bend frequencies, obtained on MJDs~53461, 53473 and 53496 respectively.  The first of these three observations was obtained at the same time as the {\it Chandra} observation which revealed a strong disc wind in this state.  As already suggested by the independence of the bend-frequency and hardness ratio, the X-ray spectra corresponding to distinct bend-frequencies are very similar, with the major difference being a small change in the normalisation and some apparent changes around the iron absorption edge that could be associated with changes in the disc wind.

Although a systematic study of the weak spectral variability in the hypersoft state is beyond the scope of this work, we have fitted these three example spectra with a simple spectral model in order to highlight the similarities and comment on possible changes throughout the hypersoft state.  We fitted a model consisting of a disc blackbody, convolved with a power-law component (using the {\sc simpl} model of \citealt{Steiner2009}) to represent inverse Compton scattering.  The resulting continuum was multiplied by a simple absorption edge to represent the iron absorption edge seen around 8~keV, together with Galactic neutral absorption (the {\sc phabs} model in {\sc xspec}) with fixed column density $N_{\rm H}=7.4\times 10^{21}$~cm$^{-2}$ \citep{Miller2008}.  The fitted model curves and data/model ratios from the fits are also shown in Fig.~\ref{hypspeccomp}. With such high signal-to-noise spectra, even including a standard systematic error of 0.5~per cent yields a poor $\chi^{2}=1183$ (for 112 degrees of freedom) for the combined fit to all three spectra. The residuals mainly appear to be systematic features which could arise because we use a simplistic single edge to model the absorption, whereas the {\it Chandra} HETG data reveal a more complex spectrum of absorption lines, as well as the expected edge.  However, a more physically self-consistent fitting of the absorption is not warranted by the {\it RXTE} PCA data which has relatively poor spectral resolution.

Detailed consideration of errors is not permitted by the overall poor fit.  However, the key robust points to note are:
\begin{enumerate}
\item The edge energy remains close to 8.3~keV, but its optical depth decreases significantly, from 0.66 in the observation with the lowest bend frequency, to 0.47 and then 0.32 for the observation with highest bend frequency.  
\item There are no systematic trends in disc blackbody inner temperature, which takes a value of $kT\simeq 1.2$~keV in all three spectra.
\item The power-law scattering fraction is consistent with being constant, at around 0.19 of the disc blackbody photons.  However, the photon index hardens from $\Gamma\simeq 6.25$, to $\Gamma\simeq 6.03$, to $\Gamma\simeq 5.62$ as the bend frequency increases.
\end{enumerate}
The small variations in the shape of the continuum (excluding edge optical depth variations) appear to be driven by changes in the spectral slope of the scattered component.  This picture is also suggested by the difference in shape between the hypersoft state spectrum and the spectrum of the first intermediate spectral state from MJD 53473 and MJD 53453 respectively, shown in Fig.~\ref{speccomp}, which have similar spectral shapes at energies below 5~keV but diverge significantly at higher energies.  Based on this qualitative spectral comparison we speculate that the hypersoft state may be more closely related to the intermediate/very-high spectral states than the soft state, except with a much steeper upscattered power-law component than is seen in the intermediate state.

\begin{figure}
\includegraphics[scale=0.33]{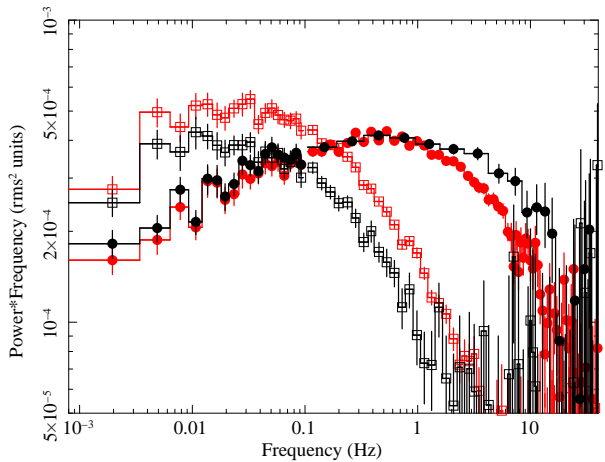}
\caption{Energy dependence of the PSD of \gro during the hypersoft state.  Red data points show the soft band PSD (2-6~keV) while black points show the PSD of the hard band (6-15~keV).  The open squares denote the PSDs of the group 2 data, while filled circles denote the PSDs of the group 6 data.}
\label{hyppsdendep}
\end{figure}

We can investigate the energy-dependence of the variability by comparing the soft-band (2-6~keV) PSDs with PSDs from the hard (6-15~keV) energy band, which are plotted in Fig.~\ref{hyppsdendep} for the group 2 and group 6 data, which have good-quality PSDs and together correspond to a significant change (factor $\sim15$) in the PSD bend frequency.  The figure confirms a trend that we see in the other data groups, namely that at low bend-frequencies the variability amplitude is larger in the soft band than the hard band (although the PSD shapes remain similar), while for PSDs with higher bend-frequencies the hard and soft band variability amplitudes are similar at low frequencies, but the hard band variability amplitude exceeds that of the soft band at high-frequencies above the bend frequency.

The overall energy-dependent differences in the PSD are small compared with the flux-dependent changes.  According to the simple disc blackbody and power-law model considered here, the upscattered power-law component contributes a significant fraction, about 20~per cent, of the observed {\it RXTE} PCA count rate in the 6-15~keV band but only $\sim6$~per cent of the count rate in the 2-6~keV band.  If the power-law alone contributed the variability, this would lead to a difference in fractional rms between the two bands of a factor $\sim 3$ and corresponding difference in PSD normalisation of a factor $\sim 10$, which is clearly not seen in the energy-dependent PSDs.  Therefore, the similarities of PSD shapes and normalisations in soft and hard bands suggest that the variability in the hypersoft state is not simply associated with the power-law tail, which appears to dominate the soft state variability in some other sources, e.g. Cyg X-1 \citep{Churazov2001}.

\section{Discussion and conclusions}
\label{discussion}
The 2005 outburst of the black hole LMXB GRO~J1655-40 was observed intensively by {\it RXTE}.  During the outburst the source showed a distinct type of soft state, lasting for $\sim40$ days and characterised by a very steep or absent power-law tail in its energy spectrum. We call this state the `hypersoft' state, due to the strong similarity of its spectrum with that of the hypersoft state first classified in the HMXB Cyg~X-3 by \citet{Koljonen2010}.  A {\it Chandra} HETG observation obtained early in the hypersoft state of GRO~J1655-40 showed evidence for a powerful equatorial disc wind, which seems to be specifically linked to this state and manifests itself in the {\it RXTE} PCA spectra as a strong edge from highly ionised iron.

We first summarise the remarkable timing and spectral properties of the hypersoft state, before comparing the behaviour with that of other timing signatures, and then going on to consider models to interpret the data:
\begin{enumerate}
\item The PSD is well-described by a bending power-law shape, with a constant low-frequency index of $-0.82$ and a steeper high-frequency index varying between -1.7 and -1.9. 
\item The PSD bend frequency shows a remarkably strong correlation with the 3--20~keV X-ray flux, with bend frequency $\nu_{\rm b}$ scaling with flux $F$ as $\nu \propto F^{8.5}$, so that the bend frequency changes by two decades during the hypersoft state while the flux varies by less than a factor of two.  There is no correlation of flux or frequency with spectral hardness.
\item A simple spectral analysis of typical {\it RXTE} PCA data from the hypersoft state shows that the spectrum can be reasonably well-modelled with a Comptonised disc blackbody with inner temperature $kT\simeq 1.2$~keV and $\sim20$~per cent of the disc photons being upscattered into a steep ($\Gamma \simeq 6$) power-law tail, which hardens only slightly as the flux and bend-frequency increase.  The optical depth of the ionised iron edge decreases significantly as the flux and bend frequency increase.
\item The energy-dependence of the PSD shows that the entire spectrum contributes to the variability, including both the disc and the upscattered component.  The energy-dependence of the PSD changes in a complex way with flux, but these are second-order effects compared with the strong flux-dependence of the bend-frequency.
\end{enumerate}
\subsection{Comparison with flux evolution of other timing signatures}
One of the legacies of the {\it Rossi X-ray Timing Explorer} has been the systematic study of X-ray timing properties in X-ray transients as they evolve throughout their outbursts.  Broadly speaking, the PSD shape is strongly linked to the X-ray spectral state, while the fractional rms amplitude is correlated with spectral hardness (e.g. \citealt{Homan2001,Belloni2005}).  At face value, there does not appear to be a strong correlation of timing properties with source luminosity, since individual states can span a wide range of luminosities.  However, if one decomposes the spectrum into disc and power-law components and focuses on individual power-spectral components, such as the low-frequency `type C' QPOs, there does appear to be a correlation between QPO frequency and disc flux, up to frequencies of $\sim5-6$~Hz \citep{Sobczak2000,Remillard2006}.  

More recent work also finds an anti-correlation between type C QPO frequency and power-law flux \citep{Motta2011}, although this anti-correlation shows multiple tracks associated with different outbursts of the same source, i.e. there is not one single relationship with power-law flux.  Interestingly, the same authors also find a positive correlation between type B QPO frequency and power-law flux.  Thus it is clear that timing signatures do evolve with the fluxes of individual spectral components, even if they do not correlate strongly with the combined flux of multiple spectral components (e.g. disc and power-law).  The spectrum of GRO~J1655-40 in its hypersoft state is probably dominated by a continuum component (perhaps a Comptonised disc blackbody) which evolves only weakly in shape or luminosity - there is no evidence for strong flux evolution of a separate spectral component which is linked to the variability.  GRO~J1655-40 shows a factor $\sim30$ increase in the bend frequency for a 50~per~cent increase in flux.  This compares with the roughly linear (or sub-linear) trends seen in variations of QPO frequencies, which lead to at most a 50~per~cent variation in frequency for the same change in flux.  Therefore, the flux-dependence of the bend frequency in the hypersoft state of GRO~J1655-40 appears to be the strongest seen in any X-ray timing component observed to date.

\subsection{Physical interpretation of the hypersoft state timing and spectral properties}
\begin{figure}
\includegraphics[scale=0.33]{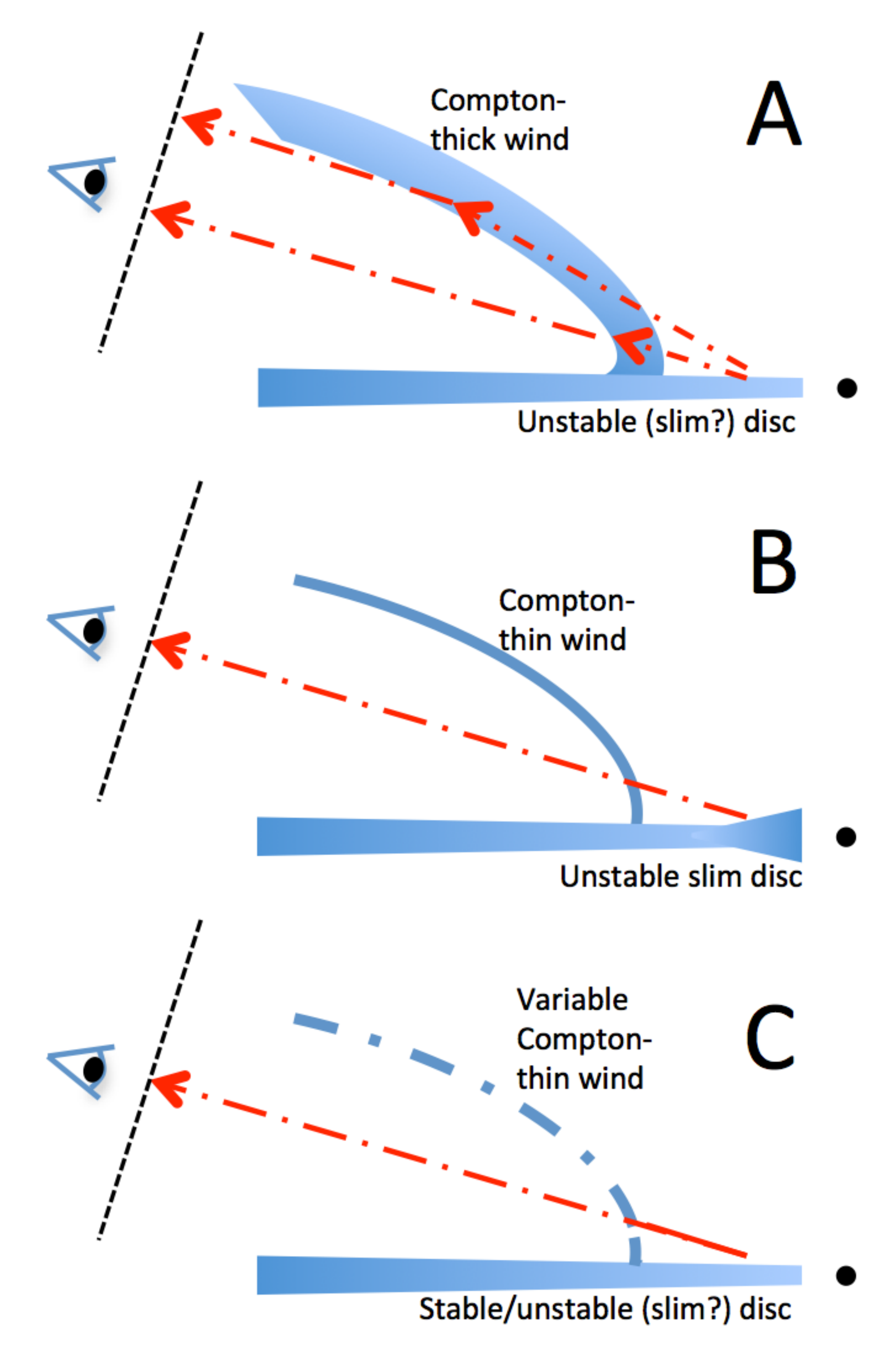}
\caption{Illustration of three different models for the timing and spectral properties of GRO~J1655-40 in the hypersoft state, showing the disc and disc-wind in blue, and photon paths in red.  The black dashed line denotes the `observer plane' (photons leaving perpendicular to this plane reach the observer).  Dot-dashed photon paths indicate variability of the photon flux (either induced by the wind or intrinsic to the disc emission itself).}
\label{modelsketch}
\end{figure}

We now consider three models to explain the unusal timing and spectral properties of the hypersoft state, which we call models A, B and C.  The models are illustrated in Fig.~\ref{modelsketch}. Since the hypersoft state in GRO~J1655-40 seems to be associated with the strong equatorial disc wind seen with the {\it Chandra} HETG instrument (and appearing as a strong iron edge in the {\it RXTE} PCA spectra), all the models considered here invoke the wind to produce the observed absorption features and in models A and C, some of the key timing properties.

\subsubsection{Model A: a wind-scattering origin for the PSD bend}
 It is natural to consider whether the strong disc wind could be responsible for the unusual spectral and timing properties of this state.  A similar approach was taken by \citet{Zdziarski2010} to the unusual spectral and timing properties of Cyg X-3, except that the wind in question in that source is assumed to be associated with the Wolf-Rayet companion star, not with the disc as must be the case in the LMXB GRO~J1655-40.  We can consider two effects linked to the scattering of X-rays by the wind.  Firstly, a sufficient column density will down-scatter the X-ray photons, removing any hard power-law tail from the energy spectrum (which may look like a more conventional intermediate state spectrum), and possibly causing the unusual spectral shape seen in the hypersoft state.  Secondly, if the wind is extended on light-second scales, the different path-lengths introduced by the scattering for light travelling from the central continuum source to the observer, will suppress the variability at high frequencies, which could explain the observed bend in the power-spectrum.

Compton down-scattering in a static optically thick medium with optical depth $\tau$, causes the spectrum to cut off at an energy $E_{\rm cut}\sim (511~{\rm keV})/\tau^2$ \citep{Sunyaev1980}.  From the observed spectra above 10~keV we should expect Compton optical depths of $\tau \sim 5$ or greater.  These values are significantly larger than inferred from the estimates of total hydrogen column density based on X-ray absorption measurements ($N_{\rm H}\sim 5\times 10^{23}\equiv \tau\sim0.3$; \citealt{Miller2008}).  However, there may be an optically thick component along the line-of-sight that is associated with the wind but does not produce absorption features, due to it being highly ionised.  

The time-scale for suppression of X-ray variations depends on the typical path-length travelled by a photon in the scattering region before it escapes, which scales with both the radius of the scattering region ($L$) and the number of scatterings experienced by the photon before it escapes (which depends on the optical depth, $\tau$).  Based on Monte Carlo simulations, \citet{Zdziarski2010} show that the intrinsic PSD of X-ray variations is damped by a factor $e$ at a frequency $\nu_{\rm scat}\simeq c/(2L\tau)$.  For comparison with the data, it is simplest to assume that $\nu_{\rm scat}$ corresponds approximately to the bend frequency in the PSD.  For GRO~J1655-40 in the hypersoft state, the large variation in bend frequency is unlikely to be associated with any significant variation in $\tau$, since the X-ray continuum is relatively constant in shape and shows no evidence for a systematic variation in cut-off energy with bend frequency.  Therefore, if scattering in the wind is the source of the bend in the PSD, the radius of the scattering region would have to vary by a factor of $>50$ to explain the observed variation in bend frequency.  Considering the very small spectral and flux changes associated with the PSD variation, it seems extremely unlikely that such a large change in wind geometry can occur during the hypersoft state and we therefore rule out scattering by the wind as the likely origin of the PSD bend.  It remains possible that the unusual spectral shape is still due to scattering by a Compton-thick component of the wind, but in this case the extent of the wind is constrained by the highest-frequency bends seen in the PSD to be significantly smaller than a light-second.

\subsubsection{Model B: variability intrinsic to the accretion flow}
On simple energetics grounds, most of the primary X-ray emission produced in the hypersoft state should originate from the inner accretion flow where most of the gravitational potential energy is released.  Viscous-time-scale variations in the accretion flow (possibly the standard disc) have been identified as a likely origin of X-ray variations on similar time-scales in other X-ray binary states \citep{Lyubarskii1997,Arevalo2006,Uttley2011}.  Therefore it is reasonable to assume that the bend in the PSD is associated with a characteristic time-scale, perhaps viscous, in the inner accretion flow of GRO~J1655-40.  The viscous time-scale scales as $t_{\rm visc}=\alpha(R/H)^{2}t_{\rm dyn}$ \citep{Treves1988} where $\alpha$ is the dimensionless viscosity parameter, $R$ and $H$ are disc radius and scale-height respectively and $t_{\rm dyn}$ is the dynamical time-scale at $R$ (scaling with $R^{3/2}$).  Variations in the PSD bend time-scale could be produced by variations in any of these parameters.  However, due to the weak spectral and flux changes it is unlikely that there are changes in a characteristic disc radius (e.g. disc truncation radius) and the parameter which could most easily explain the observed change in time-scale is the scale-height of the disc, which may be responding to changes in accretion rate.
  
In standard disc theory (e.g. see \citealt{Shakura1973}), the scale-height of gas-pressure dominated discs depends only weakly on accretion rate $\dot{M}$, scaling as $H/R \propto \dot{M}^{1/5}$ \citep{Treves1988}, which is unlikely to produce the large changes in scale-height (up to a factor $\sim10$) needed to produce the observed range of PSD bend time-scales.  However, in the radiation-pressure dominated regime $H/R \propto \dot{M}$, which could more plausibly produce the observed changes, although one must still explain the small observed luminosity change corresponding to such a relatively large change in accretion rate.  In this regime the disc must be stabilised by advection and so is more likely to be a `slim' disc \citep{Abramowicz1988} than a standard disc, which could explain the relatively small change in luminosity for a larger change in accretion rate, if most of the additional liberated gravitational potential energy is advected inwards rather than radiated.  However, such a model implies that the accretion rate in the hypersoft state is significantly larger than that inferred from the observed luminosity, since advection only affects the luminosity at close to or above the Eddington accretion rate \citep{Abramowicz1988}.  We address this issue in Section~\ref{hypersoftnature}.

\subsubsection{Model C: variability imprinted by the wind}
A remaining possibility is that the variability time-scale is intrinsic to the variability process, but that the variability is generated by variable absorption and scattering in the wind itself, e.g. the continuum source may be constant but we see variability caused by variations in column density along the line-of-sight.  Since the variations in this case would be line-of-sight only, they would be limited by the crossing time of clumps of gas in the wind across the continuum source and could be produced on relatively short time-scales, even if the wind itself is generated at relatively large radii in the disc.  For example, assuming only the $\sim400$~km~s$^{-1}$ typical radial velocity of absorption lines produced by the wind (the transverse velocity is likely to be significantly larger), a small clump of gas could cross in front of a 20~$R_{g}$ diameter emitting region in 0.5~s.  Variations of the continuum caused by Compton scattering would imply fluctuations in Compton optical depth $\Delta \tau \sim 0.1$.  Assuming there is no `hidden' component of the wind which contributes to Compton optical depth but not to the absorption features, these variations would correspond to $\sim 30$~per~cent variations in line-of-sight column density, on time-scales of seconds to minutes.

If the variability is introduced by the wind, it is unlikely that the wind is Compton-thick, because then the observed variability would be introduced at many locations in the wind (not just along the line of sight) and due to the slow wind-velocity the variations are likely to be independent from one another on the observed time-scales and hence will cancel out to produce much weaker variability.

Even if variations are imprinted on the continuum by variable line-of-sight scattering in the wind, it is still not obvious what causes the strong correlation between PSD bend-frequency and continuum flux.  It seems unlikely that such variations are produced by a changing wind geometry or velocity, for the same reasons that suppression by scattering in a wind can be ruled out as the origin of the PSD bend.  The required variations seem to be too large to be accommodated by any reasonable model.  However, it may be that the PSD shape is associated with the intrinsic accretion fluctuations which drive the wind, so that a similar explanation may be found as for the accretion flow origin of the variability (Model B) which also raises the same questions regarding the nature of the accretion flow and the underlying accretion rate in the hypersoft state.

\subsection{The nature of the hypersoft state and comparison with Cyg~X-3}
\label{hypersoftnature}
A possible explanation for the unusual nature of the hypersoft state is that the disc is accreting at a very high rate, possibly at or above the Eddington limit.  The distance to GRO~J1655-40 is not yet certain, but using the largest distance generally assumed in the literature of 3.2~kpc \citep{Hjellming1995}, the implied bolometric luminosity in the hypersoft state is around $10^{38}$~W, which is less than 20~per~cent of the Eddington luminosity for the dynamically-estimated black hole mass of 6.3~M$_{\odot}$ \citep{Greene2001} or the value of 5.31~M$_{\odot}$ estimated from QPO modelling \citep{Motta2014}.  Allowing for the high source inclination (70~degrees, \citealt{Greene2001}), the reduced solid-angle subtended by the disc as seen by the observer means that the intrinsic luminosity could be several times larger, but yields an accretion rate up to 40~per~cent of the Eddington limit, which may not be sufficient to explain the large changes in scale-height in terms of an advection-dominated slim disc model.  However, modelling of jet ejection events from GRO~J1655-40 indicates that the jet is inclined at 85~degrees to the line of sight, which may further suggest that the inner disc is more edge-on than implied by the system inclination \citep{Hjellming1995,Maccarone2002}: such a high inclination would then imply luminosities at or above the Eddington limit.

It is clearly speculative to suggest that the hypersoft state, observed in GRO~J1655-40 at an inferred bolometric luminosity which is rather typical for outbursting BHXRBs, actually corresponds to a very high accretion rate.  However, the possibility is also suggested by the unusual nature of the wind observed by {\it Chandra} during the state, which corresponds to a mass outflow rate more than a factor 10 larger than observed in the canonical soft state in the same source \citep{Ponti2012}.  The combination of an extreme accretion rate with high inclination may be what makes the hypersoft state so unusual and relatively rare, but this then leads us to question the nature of the hypersoft state in the prototypical example, Cyg~X-3.

In its hypersoft state, Cyg~X-3 shows bolometric luminosities of up to $4\times 10^{38}$~erg~s$^{-1}$ \citep{Koljonen2010}, several times larger than seen in GRO~J1655-40.  The system parameters of this Wolf-Rayet (WR) X-ray binary system are difficult to constrain and hence the nature of the compact object is unknown. However, based on evolution of the orbital period together with an estimate of the WR mass-loss rate, \citet{Zdziarski2013} inferred a low compact companion mass of $2.4^{+1.4}_{-1.0}$~M$_{\odot}$, implying an accretion rate at or exceeding the Eddington limit.  Such a high accretion rate could also be consistent with the extreme radio flux and jet ejections of Cyg X-3, which are reminiscent of those seen in the microquasar GRS~1915+105, also thought to be an extreme-accretion rate source \citep{Done2007}. It is interesting to note that Cyg X-3 becomes radio-faint in the hypersoft state \citep{Koljonen2010}.  The radio flux of GRO~J1655-40 also drops significantly during the hypersoft state, becoming undetectable\footnote{\texttt{http://www.aoc.nrao.edu/~mrupen/XRT/GRJ1655-40/\\grj1655-40.shtml}}.

The spectral similarity between GRO~J1655-40 and Cyg~X-3 in the hypersoft state suggests that the unusual spectral shape is intrinsic to this state and not due to the effects of line-of-sight scattering by the winds in each system, which have distinct origins and likely also distinct physical properties.  One clear difference between Cyg~X-3 and GRO~J1655-40 is the X-ray variability, which in Cyg~X-3 manifests above $10^{-3}$~Hz as simple red-noise with a featureless power-law PSD (with index -2) (\citealt{Axelsson2009}; we have confirmed that the same PSD is observed in the hypersoft state, which was not identified at the time of publication of that paper).  This difference might indicate that the unusual variability in GRO~J1655-40 is imprinted by the disc wind which intercepts the line-of-sight in GRO~J1655-40, but probably does not intercept the line-of-sight in Cyg~X-3, which has a lower system inclination of $\sim43$~degrees \citep{Zdziarski2013}.  However, Cyg~X-3 shows a featureless PSD regardless of spectral state, unlike GRO~J1655-40 which shows band-limited noise and QPOs in the harder states \citep{Motta2014}, suggesting that a distinct process may produce the red-noise variability in Cyg~X-3.

\subsection{Concluding remarks}
The extremely strong evolution of characteristic PSD bend-frequency with flux, with only minimal corresponding change in spectral shape, is unlike any other evolution of timing signatures seen to date in an X-ray binary. The large change in bend frequency for little spectral change suggests that the bend in the PSD is intrinsic to the variability process and cannot be caused by dilution of high-frequency variability due to scattering in the disc wind.  Possible origins of the variability are accretion variations in a high-accretion-rate slim disc, or perhaps variations in scattering column density in the wind itself, which may be driven by fluctuations in the accretion flow driving the wind.  In both these scenarios, the strong evolution of bend time-scale could be linked to changes in a fundamental parameter of the accretion flow, possibly the scale-height, since changes in the disc inner truncation radius should produce much stronger evolution of the X-ray spectrum and flux.

The discovery of the hypersoft state in GRO~J1655-40 shows that there is still much to be learned from the huge archive of spectral and timing data obtained by the {\it Rossi X-ray Timing Explorer} during its extremely productive 16 years of operations.  Regardless of the specific interpretation, the hypersoft state data suggest a tantalising link between several highly unusual phenomena: the powerful disc wind, the unusual spectral shape and the extreme evolution of timing properties.  The remarkable similarity of the spectrum with that seen from Cyg~X-3 also raises new questions about the interpretation of the hypersoft state data in that source.  Furthermore, if this intriguing new state can be linked to extreme accretion rate compact objects more generally, it could shed important light on other high-accretion-rate sources, such as high-accretion-rate AGN and any stellar-mass black holes which are observed as Ultra Luminous X-ray (ULX) sources.  For example, a recent model by \citet{Middleton2015} links the different spectral and timing properties of ULXs to the inclination-dependent effects of a powerful wind from a super-Eddington accreting black hole.  The wind in the hypersoft state of GRO~J1655-40 may play a similar role in affecting the spectral and timing properties of the source, so that the hypersoft state in XRBs may prove useful for studying the effect of winds on the behaviour of even more luminous and distant objects.

\section*{Acknowledgments}
We would like to thank Chris Done for valuable discussions and insights, and the anonymous referee for constructive comments and criticism which improved the quality of this paper.  This research has made use of data obtained from the High Energy Astrophysics Science Archive Research Center (HEASARC), provided by NASA’s Goddard Space Flight Center, and also made use of NASA’s Astrophysics Data System.

\bsp

\label{lastpage}

\end{document}